\documentclass[
journal=jacsat, 
manuscript=article]{achemso}

\usepackage[version=4]{mhchem} 

\usepackage{siunitx}
\usepackage{physics}
\usepackage{placeins}
\usepackage{amsmath}
\usepackage{amssymb}

\author{Lucas Tepper}
\affiliation[Freie Universita\"at Berlin]
{Department of Physics, Freie Universit\"at Berlin, 14195 Berlin, Germany}
\author{Benjamin Dalton}
\affiliation[Freie Universita\"at Berlin]
{Department of Physics, Freie Universit\"at Berlin, 14195 Berlin, Germany}
\author{Roland R. Netz}
\email{rnetz@physik.fu-berlin.de}
\affiliation[Freie Universita\"at Berlin]
{Department of Physics, Freie Universit\"at Berlin, 14195 Berlin, Germany}

\title[An \textsf{achemso} demo]
{Accurate Memory Kernel Extraction from Discretized Time-Series Data}

\begin{document}

\begin{abstract}
	Memory effects emerge as a fundamental consequence of dimensionality reduction when
	low-dimensional observables are used to describe the dynamics of complex many-body systems.
	In the context of molecular dynamics (MD) data analysis, accounting for memory effects
	using the framework of the generalized Langevin equation (GLE) has proven efficient, accurate and insightful, particularly when working with high-resolution
	time series data. However, in experimental systems, high-resolution data is often unavailable,
	raising questions about the impact of the data resolution on the estimated GLE parameters. This study
	demonstrates that direct memory extraction remains accurate when the discretization time is below
	the memory time. To obtain memory functions reliably even when the discretization time exceeds
	the memory time, we introduce a Gaussian Process Optimization (GPO) scheme. This scheme minimizes
	the deviation of discretized two-point correlation functions between MD and GLE simulations
	and is able to estimate accurate memory kernels as long as the discretization time stays below the longest
	time scale in the data, typically the barrier crossing time.
\end{abstract}

\section{\label{sec:Introduction}Introduction}
\begin{figure*}
	\centering
	\includegraphics[]{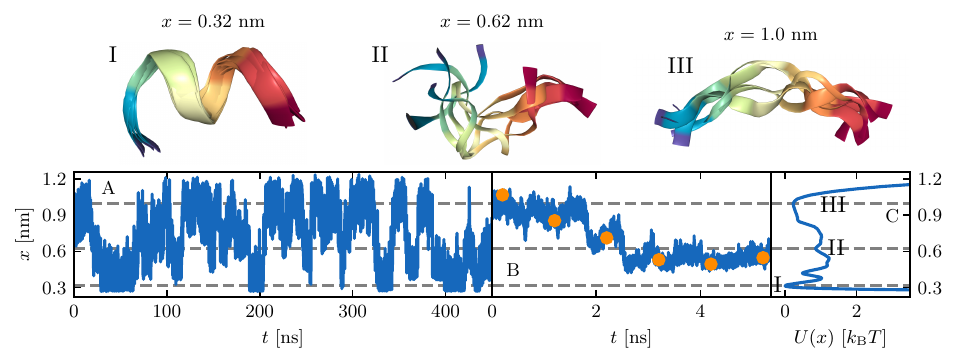}
	\caption{
		\textbf{I}-\textbf{III} Representative snapshots for different values of the mean hydrogen-bond
		distance reaction coordinate of Ala$_9$, $x$, defined in Eq.~\ref{eq:hb4}.
		\textbf{A} Multiple folding and unfolding events occur within a $\SI{450}{ns}$ trajectory segment.
		\textbf{B} A single folding event. The orange circles indicate the time series
		discretized at $\Delta t = \SI{1}{ns}$.
		\textbf{C} The potential landscape $U(x)$ for Ala$_9$, computed from the trajectory at full resolution.
		The folded state (\textbf{I})
		forms a sharp minimum at $x=\SI{0.32}{nm}$. A local minimum is found at $x=\SI{0.62}{nm}$ (\textbf{II}).
		The unfolded state forms a broad minimum around $x=\SI{1.0}{nm}$ (\textbf{III}).
	}
	\label{fig:intro}
\end{figure*}
A fundamental challenge in natural sciences involves the creation of a simplified yet
accurate representation of complex system dynamics using a low-dimensional coordinate.
For instance, in spectroscopy, atomic motions are observed solely through the polarization induced by
the electromagnetic field, resulting in spectra~\cite{quaresimaMiniReviewFunctionalNearInfrared2019}.
In the case of molecules in fluids, the myriad of interactions with the solvent are often reduced to a one-dimensional diffusion process.~\cite{einsteinUberMolekularkinetischenTheorie1905, broxOneDimensionalDiffusionProcess1986}.
In numerous studies~\cite{kitaoInvestigatingProteinDynamics1999,
	bestCoordinatedependentDiffusionProtein2010, ernstIdentificationValidationReaction2017,
	socciDiffusiveDynamicsReaction1996, neupaneProteinFoldingTrajectories2016}, the folding
of a protein is described by a one-dimensional reaction coordinate.
These diverse fields all share the common approach of projecting the complete many-body dynamics of
6N atomic positions and momenta onto a few or even a single reaction coordinate.
Starting
from the deterministic kinetics of a Hamiltonian system, the projection procedure yields a stochastic description
based on the generalized Langevin equation (GLE),~\cite{moriTransportCollectiveMotion1965,
	zwanzigNonlinearGeneralizedLangevin1973, ayazGeneralizedLangevinEquation2022}
which, in the case of a one-dimensional coordinate
$x(t)$ and its corresponding velocity $v(t)$, reads
\begin{equation}
	m \dv{t} v(t) = -\dv{U[x(t)]}{x(t)} -\int_0^t \dd s \: \Gamma(t - s) v(s) + F_R(t),
	\label{eq:gle}
\end{equation}
where $m$ is the effective mass of the coordinate $x$. The potential of mean force $U(x)$ is directly
available from the equilibrium probability distribution $\rho(x)$ via $U(x) = -k_{\mathrm{B}}T \ln \rho(x)$,
where $k_{\mathrm{B}}$ is the Boltzmann constant and $T$ the
absolute temperature. As a direct consequence of the dimensionality reduction, non-Markovian
effects arise~\cite{plotkinNonMarkovianConfigurationalDiffusion1998}.
In the GLE, the memory kernel $\Gamma(t)$ weights the effect of past velocities on the current
acceleration.
Stochastic effects, represented by the random force $F_R(t)$, are linked to the
memory function via the fluctuation-dissipation theorem in equilibrium, $\langle F_R(0) F_R(t) \rangle = k_{\mathrm{B}}T \Gamma(t)$.
When the relaxation of the environment governing $\Gamma(t)$ is sufficiently fast, $\Gamma(t)$ approaches
a delta kernel, and the Langevin equation emerges from the GLE.
Considerable efforts have been dedicated to identifying
suitable reaction coordinates to minimize memory effects and enable a Markovian description
of protein folding~\cite{bestCoordinatedependentDiffusionProtein2010,
	plotkinNonMarkovianConfigurationalDiffusion1998,neupaneProteinFoldingTrajectories2016,
	ernstIdentificationValidationReaction2017,kitaoInvestigatingProteinDynamics1999}. \\
In recent works, the memory function $\Gamma(t)$ was extracted from time series data of proteins
of biological relevance, allowing the non-Markovian description of a protein's folding kinetics in a non-linear
folding landscape. Memory effects were found to be highly relevant, both in model
systems~\cite{kapplerMemoryinducedAccelerationSlowdown2018}
and real proteins~\cite{ayazNonMarkovianModelingProtein2021,daltonFastProteinFolding2023}.
Multiple methods exist to extract memory functions from MD data.
A much-used method is based on Volterra equations, which are deterministic, integro-differential equations that allow for the
extraction of the memory kernel from time correlation
functions~\cite{berneCalculationTimeCorrelation2007, langeCollectiveLangevinDynamics2006,
	deichmannBottomupApproachRepresent2018,daldropButaneDihedralAngle2018}.
While Volterra equations offer good accuracy when high-quality time-series data is available, it is unclear
if they remain efficient when the observations of the system are sampled with long discretization times.
A recent research endeavour used an iterative scheme to approximate the memory kernel by
adapting a trial kernel with a heuristic update based on the velocity autocorrelation
function~\cite{jungIterativeReconstructionMemory2017}.
Another work parametrized memory kernels by fitting correlation functions to an
analytical solution of the GLE~\cite{daldropExternalPotentialModifies2017}.
In order to include the long-time scales of the system dynamics, the fit extended to both
the two-point correlation and its running integral.
Both methods share the limitation of not being applicable to a non-linear potential energy function $U(x)$.
A recent paper not suffering from such a limitation used a maximum-likelihood model
to estimate the GLE parameters that best fit the given MD data~\cite{vroylandtLikelihoodbasedNonMarkovianModels2022}.
In a different work on polymer solutions, star polymers were
coarse-grained to single beads interacting via a non-linear $U(x)$.
A GLE system was set up to mimic the star polymers' kinetics.
The simulation parameters of the GLE system were iteratively changed using
Gaussian Process Optimization (GPO) such that
the coarse-grained and MD velocity autocorrelations were most similar~\cite{wangDataDrivenCoarseGrainedModeling2020}.
The same idea was used to estimate a joint memory kernel over multiple
temperatures~\cite{maTransferLearningMemory2021}. \\
Here, we consider the effects of temporal discretization, motivated by the fact that data is always discretized. For
MD simulations, archived data often only contains the atomic positions at time intervals
of hundreds of picoseconds to nanoseconds, as in the case of the data from the Anton
supercomputer~\cite{lindorff-larsenHowFastFoldingProteins2011}. When considering experimental
data, measurement devices limit the time step of the observations, typically at the microsecond
scale~\cite{alemanyMechanicalFoldingUnfolding2016,vonhansenAutoCrosspowerSpectral2012}.
In a prior publication, discretization effects were examined within the framework of data-driven GLE
analysis. The GLE, without a potential, was solved analytically. To deal with discretization effects,
the discretized mean-squared displacement and velocity autocorrelation functions were computed,
allowing for the direct fitting of the memory kernel~\cite{mitterwallnerNonMarkovianDatadrivenModeling2020}.
The present work investigates how a GLE with a non-harmonic potential can be parametrized given discretized data by
considering a highly non-linear molecular dynamics test system. The Volterra-based
approach is shown to be remarkably resilient to time discretizations. Where the Volterra approach
ceases to function,
we demonstrate that Gaussian Process Optimization is a suitable method to obtain memory kernels
from discrete time series data. In matching correlation functions computed from subsampled data, we present a method
to deal with the discretization effects and extend the GLE analysis to non-linear data at higher
discretizations. The choice of correlation functions involves some flexibility, demonstrating the broad applicability
of our approach. For a small alanine homopeptide used as a test system, the Volterra method is suitable
for discretization times that reach the memory time of about $\SI{1}{ns}$. In comparison, the GPO method
extends the range to discretization times up to the folding time of about $\SI{58}{ns}$.

\section{Results and Discussion}

We investigate the effect of data discretization starting from a 10-$\mu$s-long MD trajectory of
alanine nonapeptide (Ala$_9$) in water, which was
established as a sensitive test system for non-Markovian effects in our previous work~\cite{ayazNonMarkovianModelingProtein2021}. As in our original analysis, the formation
of the $\alpha$-helix in Ala$_9$ is measured by the mean distance between the H-bond acceptor oxygen
of residue $n$ and the donor nitrogen of residue $n + 4$,
\begin{equation}
	x(t) = \frac{1}{3} \sum_{n=2}^4 || \vec{r}_n^{\,O}(t) - \vec{r}_{n + 4}^{\,N}(t) ||.
	\label{eq:hb4}
\end{equation}
In the $\alpha$-helical state, $x$ has a value
of approximately 0.3 nm, the mean H-bond length between nitrogen and oxygen. The potential of
mean force  $U(x)$ in Fig.~\ref{fig:intro}C displays several metastable states along the folding
landscape. Fig.~\ref{fig:intro}A shows a \SI{450}{ns} long trajectory.
To test how time discretization affects the memory extraction,
frames of the trajectory are left out to achieve an effective discretization time step $\Delta t$. Such a discretized
trajectory (orange data points, for $\Delta t = \SI{1}{ns}$) is compared in Fig.~\ref{fig:intro}B to the time series at full resolution. The
potential $U(x)$ is always estimated
from a histogram of the entire data to separate time discretization from effects arising due to the undersampling of the potential
(see section ~\ref{si:subsampling} in the Supporting Information).

\subsection{Volterra Equations}
\begin{figure*}
	\centering
	\includegraphics[trim={0cm 0.13cm 0 0},clip]{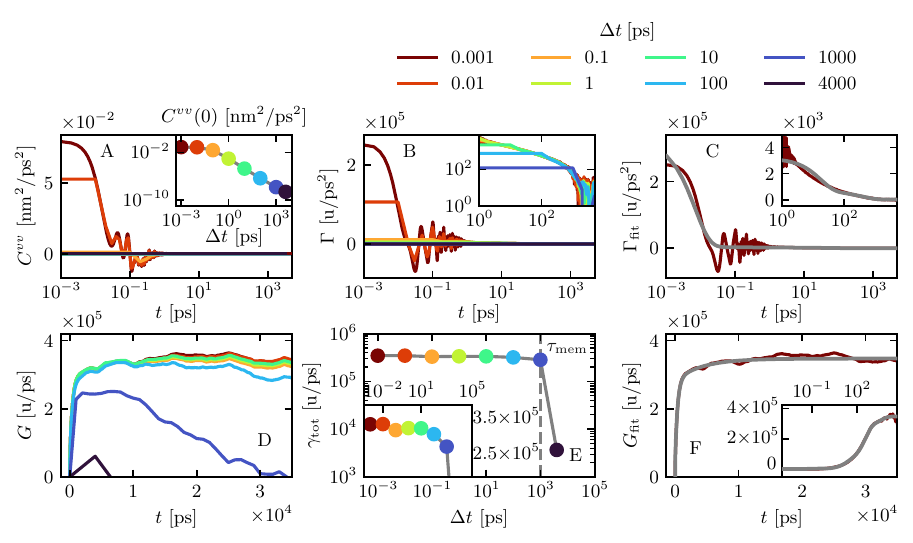}
	\includegraphics[]{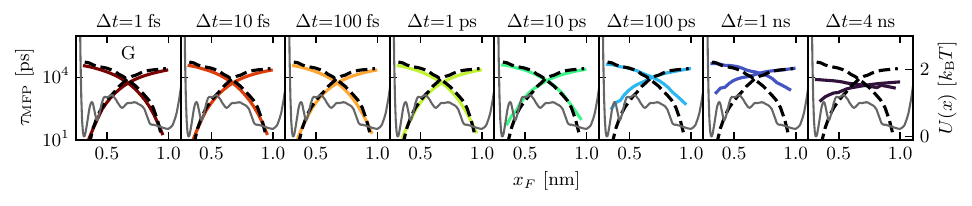}
	\caption{
		Memory extraction by the inversion of the Volterra equation~\ref{eq:volterra2} for different discretization times $\Delta t$, using MD data of Ala$_9$.
		\textbf{A}~Velocity autocorrelation $C^{vv}(t)$.
		\textbf{B}~Memory kernel $\Gamma(t)$,
		from numerical differentiation of $G(t)$.
		\textbf{C}~Multi-exponential fit of $\Gamma(t)$ computed for $\Delta t = \SI{1}{fs}$ (grey) compared to the
		numerical data (dark red). The fitted parameters are shown in Table~\ref{tab:fits_1fs} and~\ref{tab1:fits_ala9}.
		\textbf{D}~Running integral over the memory kernel $G(t)$.
		\textbf{E}~Total friction $\gamma_{\mathrm{tot}}$, computed from the exponential fits of the kernels.
		The vertical grey line indicates $\tau_{\mathrm{mem}} = \int_0^{\infty} \dd s  \: s \Gamma(s) / \int_0^{\infty} \dd s  \: \Gamma(s) = \SI{1}{ns}$.
		\textbf{F}~Fit of $G(t)$ (grey) computed at $\Delta t = \SI{1}{fs}$ compared to numerical data (dark red).
		\textbf{G}~Comparison of the mean-first passage times $\tau_{\mathrm{MFP}}$ computed from the MD data to
		$\tau_{\mathrm{MFP}}$ obtained from GLE simulations using kernels extracted at different $\Delta t$.
	}
	\label{fig:strided_ala9}
\end{figure*}

To extract memory kernels from time-series data, the  GLE in
Eq.~\ref{eq:gle} is multiplied with $v(0)$ and averaged over space or time. By using the
relation $\langle F_R(t) v(0) \rangle = 0$~\cite{moriTransportCollectiveMotion1965,zwanzigNonlinearGeneralizedLangevin1973},
one obtains the Volterra equation~\cite{daldropButaneDihedralAngle2018, ayazNonMarkovianModelingProtein2021}
\begin{equation}
	m \dv{t} C^{vv}(t) = -C^{\nabla U v}(t) - \int_0^t \dd s \Gamma(t - s) C^{vv} (s),
	\label{eq:volterra1}
\end{equation}
where $C^{vv}(t)$ is the velocity autocorrelation function and
$C^{\nabla U v}(t)$ is the correlation between the gradient of the potential and the velocity.
By integrating Eq.~\ref{eq:volterra1} from 0 to $t$, we derive a Volterra equation involving the
running integral over the kernel $G(t) = \int_0^t \dd s \Gamma(s)$
and insert~$m C^{vv}(0) = C^{\nabla U x}(0)$~\cite{ayazNonMarkovianModelingProtein2021} to obtain
\begin{equation}
	\frac{C^{\nabla U x}(0)}{C^{vv}(0)} C^{vv}(t) = C^{\nabla U x}(t) - \int_0^t \dd s G(t - s) C^{vv} (s),
	\label{eq:volterra2}
\end{equation}
with $C^{\nabla U x}(t)$ being the correlation between the gradient of the potential and the position. Computing
the memory kernel directly from Eq.~\ref{eq:volterra1} is possible~\cite{
	gordonGeneralizedLangevinModels2009, shinBrownianMotionMolecular2010}
but prone to instabilities~\cite{langeCollectiveLangevinDynamics2006}.
Extracting $G(t)$ using
Eq.~\ref{eq:volterra2} and
computing $\Gamma(t)$ via a numerical derivative improves the numerical stability~\cite{langeCollectiveLangevinDynamics2006, kowalikMemorykernelExtractionDifferent2019}.
The discretization and solution of Eq.~\ref{eq:volterra2}
is discussed in section~\ref{si:discretization} of the Supporting Information.
We fit $\Gamma(t)$ extracted from the full-resolution data at $\Delta t = \SI{1}{fs}$ using least-squares to a
multiexponential of the form
\begin{equation}
	\Gamma_{\mathrm{fit}}(t) = \sum_{i=1}^5 \frac{\gamma_i}{\tau_i} e^{-t / \tau_i}.
	\label{eq:fit}
\end{equation}
The fitted memory times $\tau_i$ and friction coefficients $\gamma_i$ are presented
in Table~\ref*{tab:fits_1fs}.
The fitting involves both $\Gamma(t)$ and $G(t)$,
as elaborated in the Methods section and accurately captures the MD kinetics, similar
to our previous work~\cite{ayazNonMarkovianModelingProtein2021}.
In order to estimate the impact of the non-Markovian effects on the kinetics,
we turn to a heuristic formula for the mean-first passage times $\tau_{\mathrm{MFP}}$ of a particle in a double well  in the presence of
exponential memory functions~\cite{kapplerMemoryinducedAccelerationSlowdown2018,kapplerNonMarkovianBarrierCrossing2019,lavacchiBarrierCrossingPresence2020}.
Validated by extensive simulations, the heuristic formula accurately described the non-Markovian effects occurring in the folding of
various proteins~\cite{daltonFastProteinFolding2023}.
For a single exponential memory function, the heuristic formula identifies
three different regimes by comparing the single memory time $\tau$ to the
diffusion time scale $\tau_D=\gamma_{\mathrm{tot}} L^2 / k_{\mathrm{B}}T$, which is the time
it takes a free Brownian particle to diffuse over a length $L$ in the reaction coordinate space.
The first regime is the Markovian limit, where $\tau \ll \tau_D$ and non-Markovian
effects are negligible. The second regime is a non-Markovian regime where $\tau_D / 100 \lesssim \tau \lesssim 10 \tau_D$,
in which a speed-up of $\tau_{\mathrm{MFP}}$ compared to the Markovian description is observed. The third regime occurs
when $\tau \gtrsim 10 \tau_D$, where $\tau_{\mathrm{MFP}}$ is slowed down compared to the Markovian description due to the
non-Markovian memory effects. \\
To compute $\tau_D$, we take $L=\SI{0.22}{nm}$, the
distance between the folded state at $x=\SI{0.32}{nm}$ and the barrier
at $x=\SI{0.54}{nm}$, the total friction $\gamma_{\mathrm{tot}} = \sum_{i=1}^5 \gamma_i$ and obtain $\tau_D = \SI{6.8}{ns}$.
The $\tau_i$ values in Table~\ref{tab:fits_1fs} span times from
$\tau_1 = \SI{7}{fs} \ll \tau_D$ up to $\tau_5 = \SI{5.7}{ns} \approx \tau_D$.
In a previous work~\cite{daltonFastProteinFolding2023},
$\tau_{\mathrm{mem}} = \int_0^{\infty} \dd s  \: s \Gamma(s) / \int_0^{\infty} \dd s  \: \Gamma(s)$, the first moment of the memory time of the kernel was proposed as the characteristic time scale
for a multi-scale memory kernel.
For the memory kernel in Table~\ref*{tab:fits_1fs}, we find $\tau_{\mathrm{mem}} = \SI{1}{ns}$, correctly predicting
the non-Markovian speed-up of $\tau_{\mathrm{MFP}}$ that a previous study demonstrated for Ala$_9$ \cite{ayazNonMarkovianModelingProtein2021}.
In this work, we will establish $\tau_{\mathrm{mem}}$ as the limit for the discretization time $\Delta t$ beyond which the
Volterra method ceases to produce accurate results. \\
\begin{table}[b]
	\centering
	\caption{
		Fitted memory function parameters for $\Delta t = \SI{1}{fs}$ according to Eq.~\ref{eq:fit}.
		The fits for $\Delta t > \SI{1}{fs}$ are shown in  section~\ref{si:fitsAla9} in the Supporting Information.
	}
	\label{tab:fits_1fs}
		\begin{tabular}{ccc}
			$i$ & $\gamma_i$ [u/ps] & $\tau_i$ [ps] \\
			\\[-1em]
			\hline
			\\[-1em]
			1 & $2.2  \cdot 10^{3}$ & 0.007 \\
			2 & $4.4  \cdot 10^{4}$ & 18 \\
			3 & $2.4  \cdot 10^{5}$ & 370 \\
			4 & $6.0  \cdot 10^{4}$ & 4100 \\
			5 & $4.6  \cdot 10^{3}$ & 5700 \\
			$\gamma_{\mathrm{tot}} = \sum_{i=1}^5 \gamma_i$ & $3.5 \cdot 10^{5}$ &  \\
			$\tau_{\mathrm{mem}} = \frac{\int_0^{\infty} \dd s  \: s \Gamma(s)}{\int_0^{\infty} \dd s  \: \Gamma(s)}$ &  & 1000 \\
		\end{tabular}
\end{table}
In the following, the full-resolution kernel obtained for a
time step of $\Delta t = \SI{1}{fs}$ will serve as a reference for results using a higher $\Delta t$.
Comparing the extracted $G(t)$ with the corresponding fit according
to Eq.~\ref{eq:fit} (red line) in Fig.~\ref{fig:strided_ala9}F
shows no significant differences in the long time limit. Fig.~\ref{fig:strided_ala9}C shows
oscillations of the extracted $\Gamma(t)$ for $t < \SI{1}{ps}$, which are discarded by the exponential fit.
As we will show later, they do not play a role in the kinetics.
For both $\Gamma(t)$ in Fig.~\ref{fig:strided_ala9}B and
$C^{vv}(t)$ in Fig.~\ref{fig:strided_ala9}A, the oscillations disappear for $\Delta t \geq \SI{0.1}{ps}$,
indicating that they are caused by sub-picosecond molecular vibrations.
Moreover, the value of $\Gamma(t)$ for $t < \SI{1}{ps}$ is consistently attenuated as $\Delta t$ increases,
mirroring the same trend observed in $C^{vv} (0)$,
as illustrated in the inset of Fig.~\ref{fig:strided_ala9}A.
In contrast, $\Gamma(t)$ for $t > \SI{1}{ps}$ in the inset of Fig.~\ref{fig:strided_ala9}B shows an exponential
decay that is well preserved for all $\Delta t < \SI{1}{ns}$. The running integral
$G(t)$ in Fig.~\ref{fig:strided_ala9}D stays mostly unchanged for discretizations smaller
than $\Delta t < \SI{1}{ns}$. This demonstrates that the Volterra extraction scheme is accurate for
discretization times below the mean memory time, i.e. for $\Delta t < \tau_{\mathrm{mem}} = \SI{1}{ns}$.\\
The multiexponential kernel in Eq.~\ref{eq:fit} allows for the efficient numerical simulation of the GLE
by setting up a Langevin equation where the reaction coordinate $x$ is coupled harmonically to one
overdamped, auxiliary variable per exponential
component~\cite{zwanzigNonlinearGeneralizedLangevin1973,baoNumericalIntegrationNonMarkovian2004}
(see section~\ref{si:markovian} in the Supporting Information).
Utilizing this simulation technique, Fig.~\ref{fig:strided_ala9}G compares profiles for the mean-first passage times $\tau_{\mathrm{MFP}}$
originating from both the folded and unfolded states.
For $\Delta t \leq \SI{10}{ps}$, the $\tau_{\mathrm{MFP}}$ values obtained from the
GLE simulations closely align with those derived from MD simulations, thereby manifesting the precise correspondence between
the non-Markovian GLE description and the kinetics observed in the MD simulation.
In Fig.~\ref{fig:strided_ala9}E, we present the asymptotic limit $\lim_{t\to\infty} G(t)$,
representing the total friction coefficient $\gamma_{\mathrm{tot}}$ of the system,
estimated by summing the individual $\gamma_i$ values obtained from the exponential fits.
When $\Delta t \geq \SI{1}{ns}$, we find that $G(t)$ does not show a plateau value in the long-time limit.
Consequently, this leads to a notable discrepancy between the $\tau_{\mathrm{MFP}}$ profiles presented in Fig.~\ref{fig:strided_ala9}G
and their MD counterparts for $\Delta t \geq \SI{1}{ns}$.
Combining the information provided in Fig.~\ref{fig:strided_ala9}D, E, and G, it becomes evident that
the extracted profile of $G(t)$, the total friction $\gamma_{\mathrm{tot}}$,
and folding times $\tau_{\mathrm{MFP}}$ all deviate significantly
from the MD reference data when the discretization time approaches the memory time $\tau_{\mathrm{mem}}$. As a result, we
assert that the Volterra extraction becomes inadequate when the discretization time exceeds the memory time $\tau_{\mathrm{mem}}$.

\subsection{Gaussian Process Optimization}
\begin{figure*}
	\centering
	\includegraphics[trim={0.05cm 0.25cm 0 0},clip]{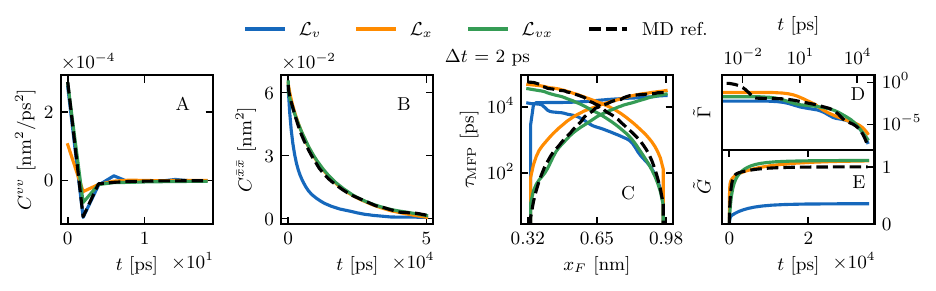}
	\includegraphics{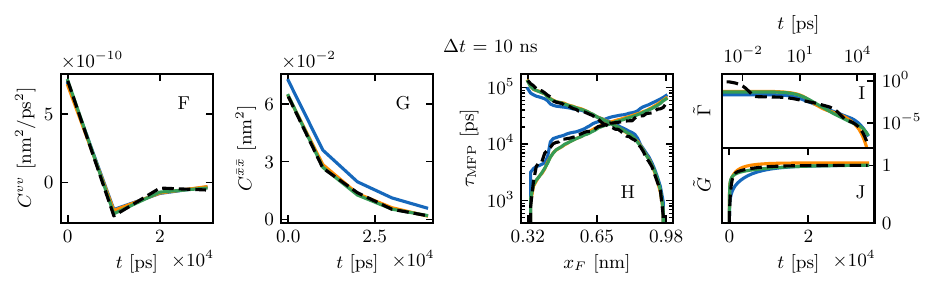}
	\caption{
		To visualize the Gaussian Process Optimization (GPO), we plot the mean of observables
		over the ten best optimization runs. We compare GPO results
		using the loss $\mathcal{L}_v$ (blue), based on
		$C^{vv}(t)$, $\mathcal{L}_x$ (orange), based on the autocorrelation of the position $\bar{x}(t) = x(t) - \langle x \rangle$
		and $\mathcal{L}_{vx}$, a linear combination of $\mathcal{L}_v$ and $\mathcal{L}_x$ (green). For $\Delta t = \SI{2}{ps}$,
		we compare the observables \textbf{A}~$C^{vv}(t)$, \textbf{B}~$C^{\bar{x} \bar{x}}(t)$, \textbf{C}~$\tau_{\mathrm{MFP}}$,
		\textbf{D}~$\tilde{\Gamma}(t) = \Gamma(t) / \Gamma(0)$ and \textbf{E}~$\tilde{G}(t) = G(t) / G(0)$
		to the MD reference (black, broken line).
		Equally, for $\Delta t = \SI{10}{ns}$, we show \textbf{F}~$C^{vv}(t)$, \textbf{G}~$C^{\bar{x} \bar{x}}(t)$, \textbf{H}~$\tau_{\mathrm{MFP}}$,
		\textbf{I}~$\tilde{\Gamma}(t)$ and \textbf{J} $\tilde{G}(t)$.
		The kernels in \textbf{D}, \textbf{E}, \textbf{I}, and \textbf{J}, parametrized by Eq~\ref{eq:fit},
		are plotted time-continuously.
	}
	\label{fig:gpo}
\end{figure*}
So far, we have demonstrated that the Volterra equation can be used to extract a consistent memory
kernel for a wide range of discretization times up to $\Delta t \approx \tau_{\mathrm{mem}}$. The resulting GLE system captures
the underlying kinetics faithfully when judged by $\tau_{\mathrm{MFP}}$ for discretizations below the
memory time scale $\tau_{\mathrm{mem}}$ but fails when exceeding it. Given that the discretization
time may exceed the dominant memory time scale in typical experimental settings, an
improved method is clearly desirable. We describe a scheme that allows the extraction of
$\Gamma(t)$ for $\Delta t$ that significantly exceeds $\tau_{\mathrm{mem}}$.
For this, we use a matching scheme between the
discretized time correlation functions of the MD reference system $C^{\textrm{MD}}(n \Delta t)$ and of
the GLE $C^{\textrm{GLE}}(n \Delta t, \theta)$ via the mean-squared loss
\begin{equation}
	\mathcal{L} = \frac{1}{N} \sum_{n = 1}^{N} \left( C^{\textrm{MD}}(n \Delta t) - C^{\textrm{GLE}}(n \Delta t, \theta) \right)^2.
	\label{eq:loss}
\end{equation}
The type of correlation function will be specified later.
The loss is evaluated over
$N$ samples, where $N$ is determined based on the decay time of the correlation (see Table~\ref{tab:results_gpo}).
In an iterative optimization,
the friction and memory time parameters in Eq.~\ref{eq:fit} that serve as the GLE parameters
$\theta = (\gamma_1, \tau_1, \ldots, \gamma_5, \tau_5)$ are updated, and the GLE
is integrated using a simulation time step $\delta t$ chosen small enough that discretization effects are negligible.
For the sake of comparability, we maintain a constant mass value of $m = \SI{31.4}{u}$,
derived using the equipartition theorem according to $m = k_{\mathrm{B}}T / \langle v^2 \rangle$,
from the MD data. In fact, the precise value of $m$ has no significant influence on
the method's outcome since it can be accommodated within the kernel. Furthermore, the system's inertial time
$\tau_{\mathrm{m}} = m / \gamma_{\mathrm{tot}} = \SI{0.09}{fs}$ is markedly shorter than all other relevant time scales, leading to an overdamped system
in which the mass value is irrelevant.
To find the best $\theta$, the choice of the optimizer is crucial. The loss $\mathcal{L}$, defined in Eq.~\ref{eq:loss}, is inherently noisy
due to the stochastic integration of the GLE and possesses, in general, many local minima in a
high-dimensional space. Faced with such a task, common gradient-based or simplex methods
fail~\cite{cetinGlobalDescentReplaces1993, bandlerOptimizationMethodsComputerAided1969}.
Genetic algorithms present a powerful alternative but require many sample evaluations~\cite{zeiglerStudyGeneticDirect1974, fitzpatrickGeneticAlgorithmsNoisy1988, bucheAcceleratingEvolutionaryAlgorithms2005}.
Given the computational cost of a converged GLE simulation, we choose Gaussian Process
Optimization (GPO)~\cite{williamGaussianProcessesMachine2006, gramacySurrogates, deringerGaussianProcessRegression2021} as a method to minimize $\mathcal{L}$.
GPO builds a surrogate model of the real
loss $\mathcal{L}$ that incorporates noise~\cite{stegleGaussianProcessRobust2008,daemiIdentificationRobustGaussian2019,lin2019robust}
and allows for non-local search~\cite{kaappaGlobalOptimizationAtomic2021,nikolaidisGaussianProcessbasedBayesian2021} (see section~\ref{si:gpo} in the Supporting Information). As an active learning technique, it guides the sampling of
new parameters, improving optimization efficiency~\cite{
	zhaoImprovingComputationalEfficiency2022,changDatadrivenExperimentalDesign2021,jinGaussianProcessassistedActive2022}. \\
\begin{figure*}
	\centering
	\includegraphics{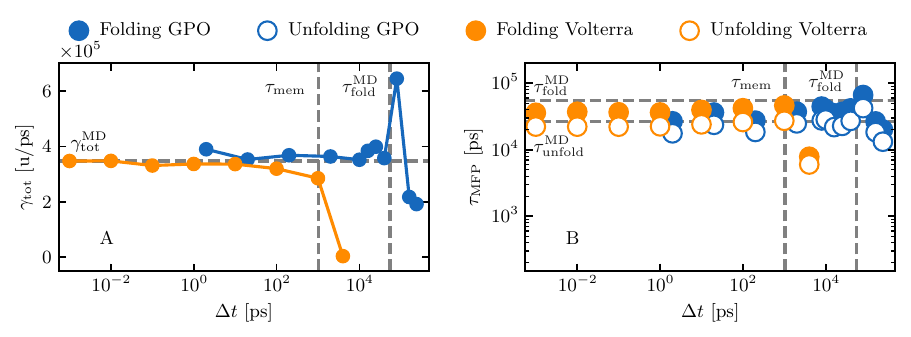}
	\caption{
		\textbf{A} The total friction $\gamma_{\mathrm{tot}} = \sum_{i=1}^5 \gamma_i$ obtained via the Volterra scheme
		(orange) is constant for discretizations of $\Delta t < \SI{1}{ns}$. For
		$\Delta t$ higher than the memory time $\tau_{\mathrm{mem}} = \SI{1}{ns}$,
		it decreases until the extraction fails.
		Gaussian process optimization (GPO, blue) estimates the correct friction
		for much higher $\Delta t$. The horizontal grey line shows $\gamma_{\mathrm{tot}}^{\mathrm{MD}}$, the total friction from the MD data.
		\textbf{B} The folding and unfolding mean-first passage times from the GLE simulations of kernels extracted
		at different discretizations, given by the mean time it takes the system
		to get from $x = \SI{0.32}{nm}$ to $x = \SI{0.98}{nm}$ and back.
		The MD folding times, $\tau^{\mathrm{MD}}_{\mathrm{fold}} = \SI{58}{ns}$ and
		$\tau^{\mathrm{MD}}_{\mathrm{unfold}} = \SI{26}{ns}$ are indicated
		as horizontal grey lines, $\tau_{\mathrm{mem}} = \SI{1}{ns}$ and $\tau^{\mathrm{MD}}_{\mathrm{fold}}$ as vertical grey lines.
		The GPO estimates the correct folding and unfolding times
		up to $\Delta t \approx \tau^{\mathrm{MD}}_{\mathrm{fold}}$,
		significantly higher than the Volterra scheme.
	}
	\label{fig:final}
\end{figure*}
In principle, any correlation function can serve as an optimization target.
Fig.~\ref{fig:strided_ala9}A shows that the velocity autocorrelation function $C^{vv}(t)$ decays to zero after about \SI{1}{ps}, while
Fig.~\ref{fig:gpo}B and G show that $C^{\bar{x} \bar{x}}(t)$,
the autocorrelation of the position $\bar{x}(t) = x(t) - \langle x \rangle$,
decays over about \SI{50}{ns}. With such a difference in the decay times of the two
correlations, we define two losses based on Eq.~\ref{eq:loss}, $\mathcal{L}_v$, using $C^{vv}(t)$, and $\mathcal{L}_x$, using $C^{\bar{x} \bar{x}}(t)$, anticipating
that the two correlations probe different scales of the dynamics. Furthermore, we define
$\mathcal{L}_{vx} = \alpha \mathcal{L}_v + \mathcal{L}_x$, a linear combination of $\mathcal{L}_v$ and $\mathcal{L}_x$, to test if including
both correlations in the loss function improves the quality of the GLE parameters.
The parameter $\alpha$ is selected for each $\Delta t$ to achieve a balanced weighting between the two losses and is tabulated in Table~\ref{tab:results_gpo} in the Supporting Information.
For every GP optimization, 300 different $\theta$ values are evaluated via a 18-$\mu$s-long
GLE simulation each. The ten
$\theta$ samples with the lowest loss form the basis for the following analysis.
When optimizing the loss function $\mathcal{L}_v$ with a discretization of $\Delta t = \SI{2}{ps}$,
Fig.~\ref{fig:gpo}A illustrates that $\mathcal{L}_v$ (blue) accurately replicates the MD reference for $C^{vv}(t)$,
whereas $\mathcal{L}_x$ (orange) exhibits discrepancies. Conversely, in Fig.~\ref{fig:gpo}B,
$\mathcal{L}_x$ perfectly reproduces $C^{\bar{x} \bar{x}}(t)$, while $\mathcal{L}_v$ struggles to do so.
Remarkably, the combined loss function $\mathcal{L}_{vx}$ (green) successfully aligns with both reference correlations simultaneously.
To evaluate the quality of the GLE parameters, Fig.~\ref{fig:gpo}C provides a comparison of the mean
first-passage times $\tau_{\mathrm{MFP}}$ between GLE results from GPO solutions and the MD reference. We calculate
$\tau_{\mathrm{MFP}}$ for GPO-based GLE simulations and the MD reference using identical discretizations $\Delta t$.
Notably, we observe that $\mathcal{L}_v$ fails to align with the MD reference, whereas both $\mathcal{L}_x$ and $\mathcal{L}_{vx}$
exhibit consistency with it. This outcome underscores the insufficiency of $C^{vv}(t)$ in
capturing the complete kinetics of the barrier crossing. A comparison of $\tau_{\mathrm{MFP}}$ between $\mathcal{L}_{vx}$ and $\mathcal{L}_x$
reveals a slightly better correspondence to the MD reference for $\mathcal{L}_{vx}$, signifying that
the inclusion of $C^{vv}(t)$ improves the optimization.
Examining the obtained kernels in Fig.~\ref{fig:gpo}D-E, all loss functions yield kernels that
largely conform to the exponential fit of the MD reference but exclude the first memory component
with a decay time of approximately $\tau_1 \approx \SI{7}{fs}$. Both $\mathcal{L}_x$ and $\mathcal{L}_{vx}$ correctly
identify the plateau of $G(t)$, while $\mathcal{L}_v$ underestimates it, leading to an underestimated $\tau_{\mathrm{MFP}}$.
Next, we evaluate the performance of the GPO for discretization times exceeding $\tau_{\mathrm{mem}}$.
In Fig.~\ref{fig:gpo}F-J, we show the results for $\Delta t = \SI{10}{ns}$, demonstrating that the GPO approach
yields similar results for all losses.
The discretized $C^{vv}(t)$, $C^{\bar{x} \bar{x}}(t)$ and $\tau_{\mathrm{MFP}}$ are in perfect
agreement with the MD reference. The kernels agree for all but the lowest times.
To confirm that the increased discretization used for the $\tau_{\mathrm{MFP}}$ computation does not introduce
any bias into the results, we perform an additional comparison of $\tau_{\mathrm{MFP}}$ computed at the full-time resolution
of $\Delta t = \SI{2}{fs}$ (see Fig.~\ref{fig:mfpts_compare_unstrided} in the Supporting Information).
Fig.~\ref{fig:final} provides a comparison of the performance of the Volterra and GPO approaches
across various discretizations. This comparison focuses on the overall friction,
folding, and unfolding mean-first passage times, as these observables are not included in the GPO optimization process.
As shown in the previous section, the applicability of the Volterra method is limited to
discretizations below the memory time $\tau_{\mathrm{mem}} = \SI{1}{ns}$.
Extraordinarily, the GPO approach can surpass the boundary set by the
memory time and estimates folding times with good accuracy for discretizations
up to $\Delta t = \SI{40}{ns}$.
This limit roughly corresponds to the mean time it takes the system to
fold, $\tau^{\mathrm{MD}}_{\mathrm{fold}} = \SI{58}{ns}$, which is given by the mean-first passage time
from the unfolded state at $x = \SI{0.98}{nm}$ to the folded state at $x = \SI{0.32}{nm}$.
For the highest discretization time tested, $\Delta t = \SI{240}{ns}$, the GP
optimization still finds meaningful folding times while underestimating the total friction.

\section{Conclusions}
We investigate the effect time discretization of the input data has on the memory extraction.
As a specific example, we consider MD time-series data of the polypeptide Ala$_9$.
Computing a memory kernel via the inversion of the Volterra equation~\ref{eq:volterra2} requires the velocity
autocorrelation and potential gradient-position correlation function. The velocity
autocorrelation changes significantly as a result of increasing time discretization, and with it,
a surrogate kernel is obtained that differs from the full-resolution kernel.
Our key finding is that given a discretization time lower than the characteristic memory time,
the Volterra approach is still able to compute a kernel that
reproduces the kinetics of the MD system.
Here, we define the characteristic memory time $\tau_{\mathrm{mem}}$ via the first moment of the memory kernel,
taking into account all decay times of the kernel and finding $\SI{1}{ns}$ for Ala$_9$.
By extracting the memory kernel from MD trajectories at different
discretizations, we show that the Volterra approach is able to reproduce the kinetics
when the discretization time $\Delta t$ is below $\tau_{\mathrm{mem}}$. \\
To also cover the important regime when $\Delta t > \tau_{\mathrm{mem}}$, we introduce a Gaussian Process Optimization scheme based on matching discretized time correlation
functions of the reference and the GLE
system.
We test losses based on the velocity and position autocorrelation functions,
for which GPO obtains memory kernels very similar to the Volterra scheme and is able to reproduce the reaction-coordinate dynamics and the folding times. \\
We demonstrate the effectiveness of GPO for discretization times up to the folding time of $\tau^{\mathrm{MD}}_{\mathrm{fold}} = \SI{58}{ns}$, about
50 times higher than the highest discretization for which the Volterra approach is applicable.
As elaborated in previous
works~\cite{kapplerMemoryinducedAccelerationSlowdown2018,ayazNonMarkovianModelingProtein2021,daltonFastProteinFolding2023},
memory can affect the kinetics of protein barrier crossing on time scales far exceeding
the memory time, up to the longest time scale of the system.
Therefore, the presented GPO approach is expected to extend the applicability of the non-Markovian analysis
to a wide range of discretized systems not suitable for the Volterra method. \\
In fact, the GPO analysis is not limited to data from MD simulations but can be used
whenever encountering highly discretized experimental data.
The application to data from single-molecule
experiments~\cite{petrosyanUnfoldedIntermediateStates2021,hinczewskiMechanicalFoldingTrajectories2013,neupaneDirectObservationTransition2016}
is a promising venue for future research.

\section{Methods}

The MD simulation data is taken from our previous publication,
see~\cite{ayazNonMarkovianModelingProtein2021} for details.
The MD simulation has a simulation time step of $\delta t = \SI{1}{fs}$,
while all GLE simulations use a time step of $\delta t = \SI{2}{fs}$.
In the computation of the hb4 coordinate (Eq.~\ref{eq:hb4}), the distances are computed between
the oxygens of Ala2, Ala3, Ala4 and the nitrogens of Ala6, Ala7, Ala8, where
Ala1 is the alanine residue at the N-terminus of the polypeptide of Ala$_9$. \\
All analysis code is written
in Python \cite{pythonmanual} or Rust \cite{matsakis2014rust}.
Table~\ref{tab:results_gpo} shows the weights $\alpha$ for the loss $\mathcal{L}_{vx}$, which includes $C^{vv}(t)$ and $C^{\bar{x} \bar{x}}(t)$.
The memory kernels are fitted using the differential evolution algorithm implemented in the Python
package 'scipy' \cite{2020SciPy-NMeth} by minimizing
a mean-squared loss, including both the kernel and the running integral over the kernel,
$\mathcal{L}_{\mathrm{mem}} = \mathcal{L}_{\mathrm{\Gamma}} + \alpha_{\mathrm{mem}} \mathcal{L}_{\mathrm{G}}$,
where $\mathcal{L}_{\mathrm{\Gamma}}$ is the mean-squared loss of the kernel and
$\mathcal{L}_{\mathrm{G}}$ is the mean-squared loss of the running integral of the kernel.
The resulting kernels and values for $\alpha_{\mathrm{mem}}$ are shown in Table~\ref{tab1:fits_ala9}. \\
The GPO is performed using the 'GaussianProcessRegressor' implemented in
the Python
package 'scikit-learn' \cite{pedregosa2011scikit}, using ten optimizer restarts. When computing the loss $\mathcal{L}$,
the correlation functions are evaluated over a finite number of sample points, $N$,
always beginning with $t = 0$. The number of sample points $N$ is given in table~\ref{tab:results_gpo}.
To minimize the expected improvement in Eq.~\ref{eq:gp_predict1} or maximize the standard deviation
in Eq. \ref{eq:gp_predict2}, we use the 'L-BFGS-B' method implemented in 'scipy' \cite{2020SciPy-NMeth}, starting from 200
random samples drawn uniformly over the space of the parameters $\theta$ (see Table~\ref{tab:bounds_gpo}).
When performing the analysis of the GPO on the basis of the ten best runs, the integrations
are repeated with a different seed for the random number generator used in the GLE integration,
ensuring that the observables are reproduced by different integration runs with
the same GLE parameters $\theta$.

\begin{suppinfo}
	Additional derivations, details for numerical implementations, detailed numerical results and additional figures.
\end{suppinfo}

\begin{acknowledgement}
	We gratefully acknowledge support by the Deutsche Forschungsgemeinschaft (DFG) Grant SFB 1449 and the European Research Council (ERC) Advanced Grant NoMaMemo No. 835117.
\end{acknowledgement}

\newpage
\section{Supporting Information}
\FloatBarrier
\subsection{\label{si:subsampling}Subsampling of the Potenital}
\begin{figure}
	\centering
	\includegraphics{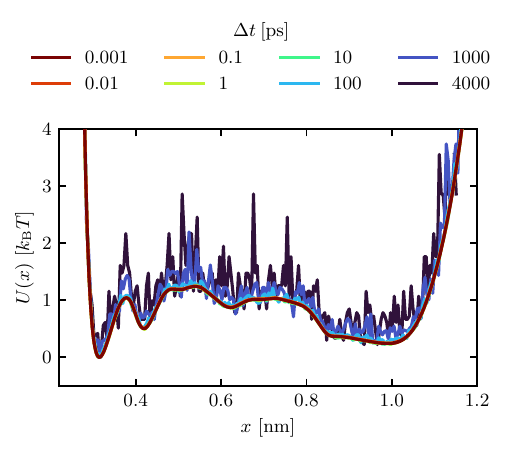}
	\caption{When generating a trajectory at an effective time step $\Delta t$ from the $\SI{10}{\mu s}$ trajectory of Ala$_9$,
		the estimated potential visibly changes for $\Delta t \geq \SI{100}{ps}$. We show the potential
		for all $\Delta t$ used in the main text. All memory extractions in the main text use the potential at full resolution, i.e.
		$\Delta t = \SI{1}{fs}$.
	}
	\label{fig:subsample}
\end{figure}

For the computation of the correlation between the gradient of the potential and the position $C^{\nabla U x}(t)$, we estimate the
density $\rho(x)$ from the trajectory via a histogram and compute the potential using $U(x) = -k_{\mathrm{B}}T \ln \rho(x)$. Fig~\ref{fig:subsample}
shows how the estimated $U(x)$ would behave if frames were left out to achieve an effective time step $\Delta t$. In the main text and all further sections, the
potential $U(x)$ is always estimated from a histogram of the entire data. Section~\ref{si:interpPot} explains how to compute $C^{\nabla U x}(t)$ from $U(x)$.

\FloatBarrier
\subsection{\label{si:discretization}Discretization of the Volterra Equation}

Starting from Eq.~\ref{eq:volterra2}, we discretize all functions of time and use the trapezoidal rule for the integral, obtaining~\cite{ayazNonMarkovianModelingProtein2021}
\begin{equation}
	G_n = \frac{2}{\Delta t C^{vv}_0} \left( C^{\nabla U x}_n - \frac{C^{\nabla U x}_0}{C^{vv}_0} C_n^{vv} -\Delta t \sum_{i = 1}^{n - 1} G_{n - i} C_i^{vv} \right).
	\label{eq:volterra_discrete3}
\end{equation}
Regarding the computation of $C^{vv}(t)$, Section~\ref{si:velocities} tests different numerical differentiation schemes to compute the velocities $v(t)$ from the positions $x(t)$.

\FloatBarrier
\subsection{\label{si:fitsAla9}Fits of the Memory Kernels}

\begin{figure}
\centering
\includegraphics{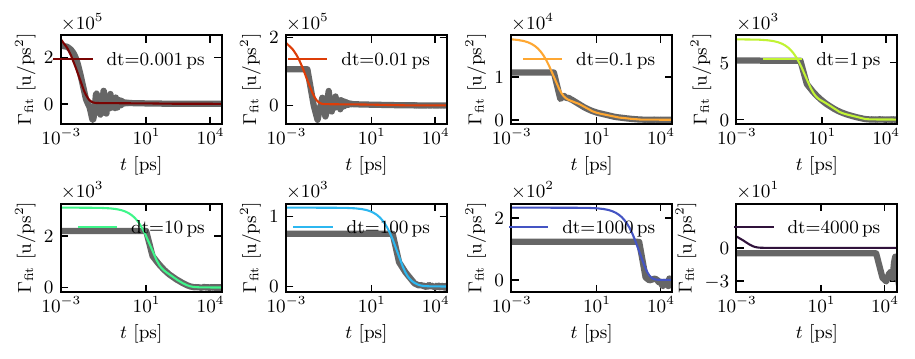}
\includegraphics{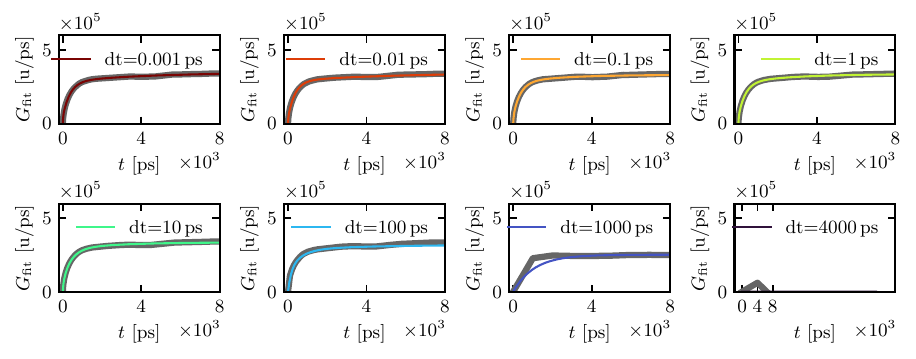}
\caption{
	Memory extraction by the inversion of the Volterra equation~\ref{eq:volterra2} for
	different discretization times $\Delta t$, using MD data of Ala$_9$. We plot the fits of
	the memory kernels shown in Table~\ref{tab1:fits_ala9}. The
	coloured lines compare the fit and the running integral
	over it to the numerical data being fitted (grey lines).
}
\label{fig:fits_ala9}
\end{figure}

We fit the memory functions for different discretizations $\Delta t$ according
to the five-component exponential fit in Eq.~\ref{eq:fit}. Table~\ref{tab1:fits_ala9}
shows the resulting memory parameters $(\gamma_1, \tau_1, \ldots \gamma_5, \tau_5)$. The fits (coloured) are compared to the memory kernels (grey) calculated using Eq.~\ref{eq:volterra_discrete3} in Fig.~\ref{fig:fits_ala9}. \\
At discretizations of $\Delta t \geq \SI{8}{ns}$, which is higher than the ones considered in this paper,
the Volterra method experiences a complete breakdown, signified by the occurrence of negative
values for $G(t > 0)$. As negative frictions are unphysical,
we do not show the corresponding $G(t)$ in Fig.~\ref{fig:fits_ala9}, nor in the main text.

\vspace{1cm}

\begin{table}
\centering
\caption{We show all memory times $\tau_i$ and
	the corresponding friction coefficients $\gamma_i$ for the five-component exponential memory kernel fit
	following Eq.~\ref{eq:fit}
	of the MD of Ala$_9$ subsampled at different time steps $\Delta t$.
	The fits are plotted in Fig.~\ref{fig:fits_ala9}.
	We also show the total friction $\gamma_{\mathrm{tot}} = \sum_{i=1}^5 \gamma_i$. The weights
	$\alpha_{\mathrm{mem}}$ are chosen such that the mean-squared loss of the kernel and the running integral
	are approximately of the same order.
	Values are shown in units of ps for $\tau_i$ and $\Delta t$, u/ps for $\gamma_i$ and atomic mass units u for $m$.
}%
\label{tab1:fits_ala9}
	\begin{tabular}{lccccccccc}
		$\Delta t$ & 0.001 & 0.01 & 0.1 & 1.0 & 10.0 & 100.0 & 1000.0 & 4000.0 \\
		\\[-1em]
		\hline
		\\[-1em]
		$\tau_1$ & 6.8e$-$03 & 9.6e$-$03 & 9.9e$-$02 & 1.7e$+$00 & 1.5e$+$01 & 1.6e$+$02 & 1.1e$+$03 & 2.5e$-$03 \\
		$\gamma_1$ & 2.2e$+$03 & 1.9e$+$03 & 1.4e$+$03 & 8.5e$+$03 & 3.3e$+$04 & 1.4e$+$05 & 2.5e$+$05 & 3.9e$-$02 \\
		$\tau_2$ & 1.8e$+$01 & 1.3e$+$01 & 2.8e$+$00 & 2.8e$+$01 & 1.0e$+$02 & 6.3e$+$02 & 5.4e$+$05 & 8.4e$+$05 \\
		$\gamma_2$ & 4.4e$+$04 & 3.6e$+$04 & 9.0e$+$03 & 4.0e$+$04 & 3.9e$+$04 & 1.5e$+$05 & 1.0e$-$04 & 5.0e$-$01 \\
		$\tau_3$ & 3.7e$+$02 & 3.5e$+$02 & 3.2e$+$01 & 3.6e$+$02 & 4.3e$+$02 & 3.7e$+$03 & 5.7e$+$05 & 9.9e$+$05 \\
		$\gamma_3$ & 2.4e$+$05 & 2.5e$+$05 & 4.0e$+$04 & 2.3e$+$05 & 2.1e$+$05 & 3.6e$+$04 & 1.0e$-$04 & 2.2e$-$01 \\
		$\tau_4$ & 4.1e$+$03 & 5.2e$+$03 & 3.6e$+$02 & 2.6e$+$03 & 1.8e$+$03 & 1.9e$+$04 & 6.2e$+$05 & 1.0e$+$06 \\
		$\gamma_4$ & 6.0e$+$04 & 5.6e$+$04 & 2.3e$+$05 & 4.6e$+$04 & 1.1e$+$02 & 0.0e$+$00 & 3.1e$+$04 & 1.3e$+$02 \\
		$\tau_5$ & 5.7e$+$03 & 8.8e$+$04 & 2.3e$+$03 & 2.8e$+$03 & 2.4e$+$03 & 4.6e$+$04 & 7.8e$+$05 & 1.0e$+$06 \\
		$\gamma_5$ & 4.5e$+$03 & 2.8e$+$03 & 5.5e$+$04 & 1.4e$+$04 & 5.1e$+$04 & 7.0e$+$02 & 2.5e$+$03 & 3.6e$+$03 \\
		$\gamma_{tot}$ & 3.5e$+$05 & 3.5e$+$05 & 3.3e$+$05 & 3.4e$+$05 & 3.4e$+$05 & 3.2e$+$05 & 2.8e$+$05 & 3.7e$+$03 \\
		$\alpha_{\mathrm{mem}}$ & 1.0e$-$01 & 5.0e$-$03 & 1.0e$-$04 & 1.0e$-$05 & 5.0e$-$06 & 1.0e$-$09 & 1.0e$-$11 & 1.0e$-$14 \\
		$m$ & 3.1e$+$01 & 3.3e$+$01 & 1.0e$+$02 & 2.7e$+$03 & 1.2e$+$05 & 5.2e$+$06 & 1.6e$+$08 & 9.6e$+$08 \\
	\end{tabular}
\end{table}

\FloatBarrier
\subsection{\label{si:markovian}Markovian Embedding}

The generalized Langevin equation \\
\begin{equation}
	m \dot{v}(t) = -\nabla U[x(t)] - \int_{t_0}^t \dd s \: {\Gamma(t - s) v(t)} + F_R(t),
	\label{eq:gleSI}
\end{equation}
can be simulated by the following set of coupled Langevin equations
\begin{align}
	\label{eq:mark1}
	m \dot{v}(t) &= -\nabla U[x(t)] + \sum_{i = 1}^{5} \frac{\gamma_i}{\tau_i}  \left( y_i(t) - x(t) \right), \\
	y_i(t) &= - \frac{1}{\tau_i}  \left( y_i(t) - x(t) \right)
	+ \frac{\eta_i(t)}{\gamma_i},
	\label{eq:mark2}
\end{align}
where the random forces $\eta_{i}$ fulfill the fluctuation-dissipation theorem \\
$\langle \eta_{i}(t) \eta_{j}(t') \rangle = 2 k_{\mathrm{B}}T \gamma_{i} \delta_{ij} \delta(t - t')$.
The form of $\Gamma(t)$ is given by
\begin{equation}
	\Gamma(t) = \sum_{i=1}^{5} \frac{\gamma_i}{\tau_i} \exp\left(-\frac{t}{\tau_i} \right).
	\label{eq:kernel_gamm}
\end{equation}
Solving for $y_i(t)$ in Eq.~\ref{eq:mark2}
and inserting into Eq.~\ref{eq:mark1} yields the solution
\begin{equation}
	\begin{split}
		m \dot{v}(t) &= -\nabla U[x(t)] - \int_{t_0}^t \dd s  \: \sum_{i = 1}^{5} \frac{\gamma_i}{\tau_i} e^{-(t - s)/\tau_i} v(s) \\
		&+ \int_{t_0}^t \dd s  \: \sum_{i = 1}^{5} e^{-(t - s)/\tau_i} \frac{\eta_i(s)}{\tau_i} \\
		&+ \sum_{i = 1}^{5} \frac{\gamma_i}{\tau_i} e^{-(t - t_0)/\tau_i} \left( y_{i}(t_0) - x(t_0) \right),
	\end{split}
	\label{eqmark:result}
\end{equation}
which is identical to Eq.~\ref{eq:gleSI} with a random force which satisfies the fluctuation-dissipation theorem
$\langle F_R(0) F_R(t) \rangle = k_{\mathrm{B}}T \Gamma(t)$. The mass $m$ is computed via the equipartition theorem $m = k_{\mathrm{B}}T / \langle v^2 \rangle$.

\FloatBarrier
\subsection{\label{si:gpo}Gaussian Process Optimization}

\begin{table}
	\centering
	\caption{Bounds $B$ for the individual parameters of $\theta = (\gamma_1, \tau_1, \ldots, \gamma_5, \tau_5)$ in the GPO.}%
	\label{tab:bounds_gpo}
	\begin{tabular}{lcccccccccc}
		& $B_{\tau_1}$ & $B_{\tau_2}$ & $B_{\tau_3}$ & $B_{\tau_4}$ & $B_{\tau_5}$ & $B_{\gamma_1}$ & $B_{\gamma_2}$ & $B_{\gamma_3}$ & $B_{\gamma_4}$ & $B_{\gamma_5}$ \\
		\\[-1em]
		\hline
		\\[-1em]
		$B_{\mathrm{lower}}$ & 0.9 & 5 & 10 & 30 & 30 & 10 & 10 & 10 & 10 & 10 \\
		$B_{\mathrm{upper}}$ & $1 \cdot 10^{2}$ & $1 \cdot 10^{3}$ & $1 \cdot 10^{4}$ & $1 \cdot 10^{4}$ & $1 \cdot 10^{4}$
		& $1 \cdot 10^{4}$ & $2 \cdot 10^{5}$ & $4 \cdot 10^{5}$ & $6 \cdot 10^{5}$ & $6 \cdot 10^{5}$ \\
	\end{tabular}
\end{table}

\begin{figure}
	\centering
	\includegraphics{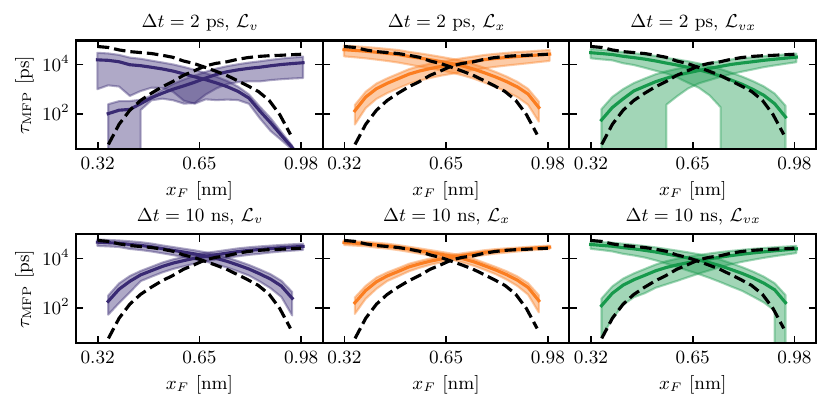}
	\caption{Comparison of $\tau_{\mathrm{MFP}}$
	starting from the folded state at $x_S = \SI{0.32}{nm}$
		and the unfolded state at $x_S = \SI{0.98}{nm}$ as a function of the endpoint of $\tau_{\mathrm{MFP}}$,
		$x_F$, for GPO runs using $\mathcal{L}_v$, $\mathcal{L}_x$ and $\mathcal{L}_{vx}$ at $\Delta t = \SI{2}{ps}$ and $\Delta t = \SI{10}{ns}$.
		We show the mean of the ten best GPO runs (coloured lines) and the standard deviation
		over the ten best runs (shaded area).
		Here, $\tau_{\mathrm{MFP}}$
		is computed from full-time resolution data, i.e. $\Delta t = \SI{2}{fs}$.
			The MD reference is indicated by the black-broken line.
	}
	\label{fig:mfpts_compare_unstrided}
\end{figure}

GPO is a widely known method, which will be briefly discussed here. For a more thorough
explanation, the reader is referred to Ref~\cite{williamGaussianProcessesMachine2006}.
The quality of a set of GLE parameters $\theta = (\gamma_, \tau_1, \ldots, \gamma_5, \tau_5)$
over the parameter space given in Table~\ref{tab:bounds_gpo} is characterized by the loss
$\mathcal{L}$ in Eq.~\ref{eq:loss} via a GLE simulation. To enhance the efficiency of the GPO for a wide range of the input parameters $\theta$, we use a logarithmic embedding
$\tilde{\theta} = (\log_{10} \gamma_1, \log_{10} \tau_1, \ldots, \log_{10} \gamma_i, \log_{10} \tau_i)$.
We assume that for a set of $n$
simulation parameters $\mathbf{\tilde{\Theta}} = [\tilde{\theta}_1, \ldots, \tilde{\theta}_n]^T$ with results
$\mathcal{L}(\mathbf{\tilde{\Theta}})$ = $[\mathcal{L}(\tilde{\theta}_1), \ldots, \mathcal{L}(\tilde{\theta}_n)]^T$,
the loss follows a Gaussian distribution:
\begin{equation}
	p
	\left(
	\begin{bmatrix}
		\mathcal{L}(\tilde{\theta}_1) \\ \ldots \\ \mathcal{L}(\tilde{\theta}_n)
	\end{bmatrix}
	\right)
	= \mathcal{N}
	\left(
	\begin{bmatrix}
		\mu(\tilde{\theta}_1) \\ \ldots \\ \mu(\tilde{\theta}_n)
	\end{bmatrix},
	\begin{bmatrix}
		k(\tilde{\theta}_1,\tilde{\theta}_1) & \ldots & k(\tilde{\theta}_1,\tilde{\theta}_n) \\
		\ldots & \ldots & \ldots \\
		k(\tilde{\theta}_n,\tilde{\theta}_1) & \ldots & k(\tilde{\theta}_n,\tilde{\theta}_n)
	\end{bmatrix}
	\right),
\end{equation}
where $\mathcal{N}(\mu, \sigma)$ is a multivariate normal distribution with mean vector
$\mu$ and covariance matrix $\sigma$. The process generating the set $\mathbf{\tilde{\Theta}}$ is a Gaussian process with the mean function
$\mu(\tilde{\theta}_i)$ and the covariance function $k(\tilde{\theta}_i, \tilde{\theta}_j)$. The mean function is set to the scalar
value $\mu(\tilde{\theta}_i) = \frac{1}{n} \sum_{i=1}^n \mathcal{L}(\tilde{\theta}_i)$. The log-likelihood
follows as
\begin{equation}
	\begin{split}
		\log p(\mathcal{L}(\mathbf{\tilde{\Theta}})) &= - \frac{1}{2} ( N \log 2 \pi + \log
		\det(k \left( \mathbf{\tilde{\Theta}}, \mathbf{\tilde{\Theta}} \right)) \\
		&+ (\mathbf{\tilde{\Theta}} - \mu)^T k \left( \mathbf{\tilde{\Theta}}, \mathbf{\tilde{\Theta}} \right) (\mathbf{\tilde{\Theta}} - \mu)
		),
		\label{eq:ll}
	\end{split}
\end{equation}
where $N$ is the dimension of $\mathbf{\tilde{\Theta}}$.
To better model data containing noisy samples, a scalar noise parameter $\sigma$ is added to the
diagonal elements of the covariance matrix
\begin{equation}
	\begin{split}
		\log p(\mathcal{L}(\mathbf{\tilde{\Theta}})) &=
		- \frac{1}{2} ( N \log 2 \pi + \log \det(k ( \mathbf{\tilde{\Theta}}, \mathbf{\tilde{\Theta}} ) + \sigma^2 I) \\
		&+ (\mathbf{\tilde{\Theta}} - \mu)^T (k (\mathbf{\tilde{\Theta}}, \mathbf{\tilde{\Theta}})+ \sigma^2 I) (\mathbf{\tilde{\Theta}} - \mu)
		),
		\label{eq:ll_noise}
	\end{split}
\end{equation}
where $I$ is the identity matrix. The complexity of the loss function the process can represent is determined by the choice of $k(\tilde{\theta}_i, \tilde{\theta}_j)$.
For $k(\tilde{\theta}_i, \tilde{\theta}_j)$, we choose the radial basis function (RBF) covariance function with an added constant $c$
\begin{equation}
	k(\tilde{\theta}_i, \tilde{\theta}_j, s, l, c) = s^2 \exp\left(-\frac{||\tilde{\theta}_i - \tilde{\theta}_j ||^2}{2 l^2} \right) + c,
	\label{eq:rbf}
\end{equation}
where $|| \cdot ||$ is the Euclidean distance. To determine the parameters $s$, $l$ and $c$, the negative log-likelihood
in Eq.~\ref{eq:ll} of the observed loss $\log \mathcal{L}(\mathbf{\tilde{\Theta}})$ is minimized, i.e. $\mathrm{arg}\,\mathrm{min}_{s, l, c} -\log p(\mathcal{L}(\mathbf{\tilde{\Theta}}))$.
The noise level $\sigma$ is kept constant at 0.005.
\\
To predict the loss value at samples not observed so far,
the data vector $\mathbf{\tilde{\Theta}}$ is split into the observed samples $\mathbf{\tilde{\Theta}}$ (where $\mathcal{L}(\mathbf{\tilde{\Theta}})$
is available) and unobserved samples $\mathbf{\tilde{\Theta}}_*$. The resulting covariance matrix has four blocks,
the covariance of $\mathbf{\tilde{\Theta}}$ and $\mathbf{\tilde{\Theta}}_*$ with themselves and each other. The distribution

\begin{equation}
	p \left( \left[ \begin{array}{c}
		\mathcal{L}(\mathbf{\tilde{\Theta}}) \\
		\mathcal{L}(\mathbf{\tilde{\Theta}}_{*})
	\end{array}
	\right] \right) =
	\mathcal{N} \left(
	\left[ \begin{array}{c}
		\mu \\ \mu
	\end{array} \right],
	\left[ \begin{array}{ll}
		k(\mathbf{\tilde{\Theta}}, \mathbf{\tilde{\Theta}}) + \sigma^2 I & k(\mathbf{\tilde{\Theta}}, \mathbf{\tilde{\Theta}}_{*}) \\
		k(\mathbf{\tilde{\Theta}}, \mathbf{\tilde{\Theta}}_{*})^T & k(\mathbf{\tilde{\Theta}}_{*}, \mathbf{\tilde{\Theta}}_{*}) \\
	\end{array}
	\right]
	\right),
\end{equation}

can be conditioned on the samples observed $\mathbf{\tilde{\Theta}}$ and their losses $\mathcal{L}(\mathbf{\tilde{\Theta}})$,
yielding a normal distribution over $\mathcal{L}(\mathbf{\tilde{\Theta}}_*)$ with mean
\begin{equation}
	M\left[ \mathcal{L}(\mathbf{\tilde{\Theta}}_*) \right] = k(\mathbf{\tilde{\Theta}}, \mathbf{\tilde{\Theta}}_*)
	\left[ k(\mathbf{\tilde{\Theta}}, \mathbf{\tilde{\Theta}}) + \sigma^2 I \right]^{-1}  \mathcal{L}(\mathbf{\tilde{\Theta}}),
	\label{eq:gp_predict1}
\end{equation}
and standard deviation
\begin{equation}
	\begin{split}
		\Sigma\left[ \mathcal{L}(\mathbf{\tilde{\Theta}}_*) \right] &=
		k(\mathbf{\tilde{\Theta}}_*, \mathbf{\tilde{\Theta}}_*) \\
		&- k(\mathbf{\tilde{\Theta}}, \mathbf{\tilde{\Theta}}_*)^T
		\left[ k(\mathbf{\tilde{\Theta}}, \mathbf{\tilde{\Theta}}) + \sigma^2 I \right]^{-1}
		k(\mathbf{\tilde{\Theta}}, \mathbf{\tilde{\Theta}}_*).
		\label{eq:gp_predict2}
	\end{split}
\end{equation}
To propose new samples, the loss space is explored or exploited. When exploiting, we
maximize the expected improvement
\begin{equation}
	\begin{split}
		EI(\tilde{\theta}_i) &= \mathbb{E}\left[ \mathcal{L}_\mathrm{best} + \xi - M[\mathcal{L}(\tilde{\theta}_i)] \right], \\
		&= \left( \mathcal{L}_\mathrm{best} + \xi - M[\mathcal{L}(\tilde{\theta}_i)] \right)
		\Phi \left( \frac{\mathcal{L}_\mathrm{best} + \xi - M[\mathcal{L}(\tilde{\theta}_i)]}{\Sigma[ \tilde{\theta}_i ]} \right) \\
		&+ \Sigma[ \tilde{\theta}_i ] \phi \left( \frac{\mathcal{L}_\mathrm{best} + \xi - M[\mathcal{L}(\tilde{\theta}_i)]}{\Sigma[ \tilde{\theta}_i ]} \right),
	\end{split}
\end{equation}
where $\mathcal{L}_{\text{best}}$ is the best loss observed so far, $\phi$ is the probability density
function and $\Phi$ the cumulative distribution function
of the standard normal distribution. To encourage sampling around the $\tilde{\theta}$ with the best loss,
we smooth the expected improvement by adding the parameter $\xi = 0.05$. \\
When exploiting, we  maximize the standard deviation of the Gaussian Process
$\Sigma\left[ \mathcal{L}(\tilde{\theta}_i) \right]$ in Eq.~\ref{eq:gp_predict2}. In each GPO run, we begin
with five samples drawn uniformly over the sample space given by the bounds in Table~\ref{tab:bounds_gpo},
then we explore for 25 iterations, followed by 270 iterations
where we alternatingly explore and exploit. \\
Table~\ref{tab:results_gpo} shows all results of the GP optimizations that are shown in
Fig.~\ref{fig:gpo} and~\ref{fig:final} and discussed in the main text. Fig.~\ref{fig:mfpts_compare_unstrided}
repeats the mean-first passage time calculations from Fig~\ref{fig:gpo}, while computing $\tau_{\mathrm{MFP}}$
not using the discretization time $\Delta t$ of the GP optimization but the full resolution
of the GLE simulation, i.e. $\Delta t = \SI{2}{fs}$

\begin{table}
	\centering
	\caption{GPO results from Fig~\ref{fig:gpo} and Fig~\ref{fig:final}.
		 We show the sum of the friction,
		$\gamma_{\mathrm{tot}} = \sum_{i = 1}^5 \gamma_i$, the mean first passage time
		from $x = \SI{0.98}{nm}$ to $x = \SI{0.32}{nm}$,
		$\tau_{\mathrm{fold}}$, the mean first passage time from $x = \SI{0.32}{nm}$ to $x = \SI{0.98}{nm}$,
		$\tau_{\mathrm{unfold}}$, and the weights $\alpha$ for
		$\mathcal{L}_{vx}$ with $\mathcal{L}_{vx} = \alpha \mathcal{L}_v + \mathcal{L}_x$.
		The folding and unfolding times are not comparable when computed from the
		data at different discretizations, so we compare $\tau_{\mathrm{MFP}}$
		computed from a GLE simulation trajectory
		at the full resolution of $\Delta t = \SI{2}{fs}$.
		As in the main text, $\tau_{\mathrm{fold}}$ and $\tau_{\mathrm{unfold}}$ are computed via a mean over the ten best GPO runs.
		$\alpha$ is determined such that the
		magnitude of $\mathcal{L}_v$ and $\mathcal{L}_x$ is roughly equal. The table also shows the number of
		samples $N_v$ and $N_x$ used for the GPO losses, as given in Eq.~\ref{eq:loss}.
	}
	\label{tab:results_gpo}
	\begin{tabular}{cccccccc}
		$\mathcal{L}$ & $\Delta t [\mathrm{ps}]$ & $\gamma_{\mathrm{tot}} [\mathrm{u}/\mathrm{ps}]$ & $\tau_{\mathrm{fold}} [\mathrm{ps}]$ & $\tau_{\mathrm{unfold}} [\mathrm{ps}]$ & $\alpha$ & $N_{v}$ & $N_{x}$ \\
		\\[-1em]
		\hline
		\\[-1em]
		$\mathcal{L}_{v}$ & $2.0 \cdot 10^0$ & $1.2 \cdot 10^5$ & $9.8 \cdot 10^3$ & $8.1 \cdot 10^3$ & - & $1.0 \cdot 10^1$ & - \\
		$\mathcal{L}_{x}$ & $2.0 \cdot 10^0$ & $3.9 \cdot 10^5$ & $4.6 \cdot 10^4$ & $2.9 \cdot 10^4$ & - & - & $2.5 \cdot 10^4$ \\
		$\mathcal{L}_{vx}$ & $2.0 \cdot 10^0$ & $3.9 \cdot 10^5$ & $2.4 \cdot 10^4$ & $1.6 \cdot 10^4$ & $5.0 \cdot 10^3$ & $1.0 \cdot 10^1$ & $2.5 \cdot 10^4$ \\
		$\mathcal{L}_{vx}$ & $2.0 \cdot 10^1$ & $3.5 \cdot 10^5$ & $3.6 \cdot 10^4$ & $2.4 \cdot 10^4$ & $1.0 \cdot 10^5$ & $1.0 \cdot 10^1$ & $2.5 \cdot 10^3$ \\
		$\mathcal{L}_{vx}$ & $2.0 \cdot 10^2$ & $3.7 \cdot 10^5$ & $2.7 \cdot 10^4$ & $1.9 \cdot 10^4$ & $1.0 \cdot 10^5$ & $1.0 \cdot 10^1$ & $2.5 \cdot 10^2$ \\
		$\mathcal{L}_{vx}$ & $2.0 \cdot 10^3$ & $3.6 \cdot 10^5$ & $3.7 \cdot 10^4$ & $2.5 \cdot 10^4$ & $1.0 \cdot 10^4$ & $1.0 \cdot 10^1$ & $2.5 \cdot 10^1$ \\
		$\mathcal{L}_{vx}$ & $8.0 \cdot 10^3$ & $3.8 \cdot 10^5$ & $4.5 \cdot 10^4$ & $2.8 \cdot 10^4$ & $1.0 \cdot 10^5$ & $4.0 \cdot 10^0$ & $6.0 \cdot 10^0$ \\
		$\mathcal{L}_{v}$ & $1.0 \cdot 10^4$ & $3.5 \cdot 10^5$ & $3.8 \cdot 10^4$ & $3.0 \cdot 10^4$ & - & $4.0 \cdot 10^0$ & - \\
		$\mathcal{L}_{x}$ & $1.0 \cdot 10^4$ & $3.6 \cdot 10^5$ & $4.2 \cdot 10^4$ & $2.8 \cdot 10^4$ & - & - & $5.0 \cdot 10^0$ \\
		$\mathcal{L}_{vx}$ & $1.0 \cdot 10^4$ & $3.5 \cdot 10^5$ & $4.2 \cdot 10^4$ & $2.8 \cdot 10^4$ & $1.0 \cdot 10^5$ & $4.0 \cdot 10^0$ & $5.0 \cdot 10^0$ \\
		$\mathcal{L}_{vx}$ & $1.6 \cdot 10^4$ & $3.8 \cdot 10^5$ & $3.5 \cdot 10^4$ & $2.2 \cdot 10^4$ & $1.0 \cdot 10^5$ & $4.0 \cdot 10^0$ & $4.0 \cdot 10^0$ \\
		$\mathcal{L}_{vx}$ & $2.5 \cdot 10^4$ & $4.0 \cdot 10^5$ & $3.7 \cdot 10^4$ & $2.3 \cdot 10^4$ & $1.0 \cdot 10^5$ & $4.0 \cdot 10^0$ & $4.0 \cdot 10^0$ \\
		$\mathcal{L}_{vx}$ & $4.0 \cdot 10^4$ & $3.6 \cdot 10^5$ & $4.2 \cdot 10^4$ & $2.7 \cdot 10^4$ & $1.0 \cdot 10^5$ & $4.0 \cdot 10^0$ & $4.0 \cdot 10^0$ \\
		$\mathcal{L}_{vx}$ & $8.0 \cdot 10^4$ & $6.4 \cdot 10^5$ & $6.7 \cdot 10^4$ & $4.2 \cdot 10^4$ & $1.0 \cdot 10^5$ & $4.0 \cdot 10^0$ & $4.0 \cdot 10^0$ \\
		$\mathcal{L}_{vx}$ & $1.6 \cdot 10^5$ & $2.2 \cdot 10^5$ & $2.7 \cdot 10^4$ & $1.8 \cdot 10^4$ & $1.0 \cdot 10^5$ & $4.0 \cdot 10^0$ & $4.0 \cdot 10^0$ \\
		$\mathcal{L}_{vx}$ & $2.4 \cdot 10^5$ & $1.9 \cdot 10^5$ & $2.0 \cdot 10^4$ & $1.3 \cdot 10^4$ & $1.0 \cdot 10^5$ & $4.0 \cdot 10^0$ & $4.0 \cdot 10^0$ \\
	\end{tabular}
\end{table}

\FloatBarrier
\subsection{\label{si:interpPot}Interpolation of the Potenital}
\begin{figure}
	\centering
	\includegraphics[width=0.72\textwidth]{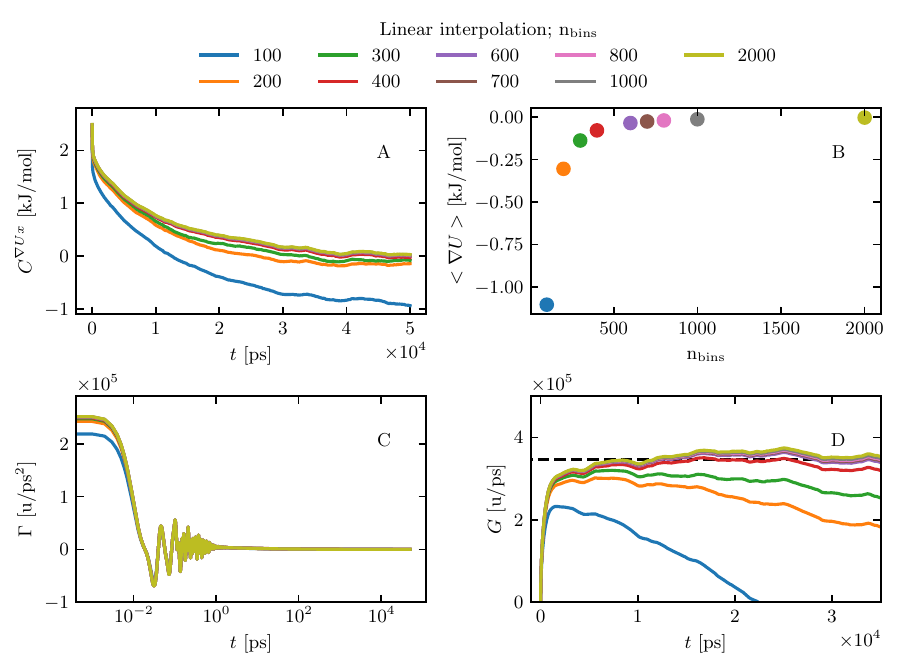}
	\includegraphics[width=0.72\textwidth]{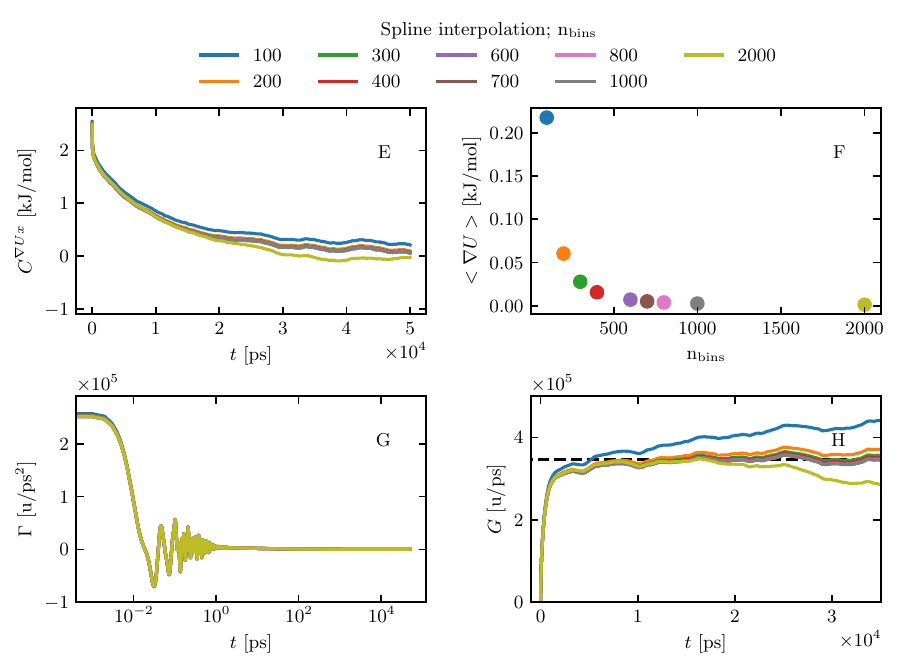}
	\caption{
		Memory extraction by the inversion of the Volterra equation~\ref{eq:volterra2} using MD data of Ala$_9$. We compare how the correlation between the gradient of the potential and the position $C^{\nabla U x}(t)$,
		the kernel $\Gamma(t)$ and the running integral of the kernel $G(t)$ depend on the number of
		interpolation points $n_{\mathrm{bins}}$ and interpolation type of the potential. The
		dependence on $n_{\mathrm{bins}}$ vanishes when $\langle \nabla U \rangle $ is close to zero.
		We show $C^{\nabla U x}(t)$ (\textbf{A}), $\langle \nabla U \rangle $ (\textbf{B}), $\Gamma(t)$ (\textbf{C}) and $G(t)$ (\textbf{D})
		for different $n_{\mathrm{bins}}$ using a linear interpolation. We show the same for a cubic spline interpolation, i.e.
		$C^{\nabla U x}(t)$ (\textbf{E}), $\langle \nabla U \rangle $ (\textbf{F}), $\Gamma(t)$ (\textbf{G}) and $G(t)$ (\textbf{H}).
		In \textbf{D} and \textbf{H}, the black, dashed line indicates $\gamma_{\mathrm{tot}} = \sum_i \gamma_i$
		from Table~\ref{tab:fits_1fs}.
	}
	\label{fig:compare_nbins}
\end{figure}

To compute the correlation between the gradient of the potential and the position $C^{\nabla U x}(t)$, we first
estimate the probability $\rho(x)$ via a histogram using $n_{\mathrm{bins}}$. Next,
we compute the potential $U(x) = -k_{\mathrm{B}}T \ln \rho(x)$. To compute a trajectory $\nabla U[x(t)]$, we
compute a numerical gradient of $U(x)$ and interpolate it such that it can be evaluated at every
frame $x(t)$. Finally, we correlate the trajectories $\nabla U[x(t)]$ and $x(t)$, yielding $C^{\nabla U x}(t)$.
Here, we compare how $C^{\nabla U x}(t)$ and $G(t)$ are influenced by the choice of $n_{\mathrm{bins}}$. We
also compare a linear to a cubic spline interpolation. Fig.~\ref{fig:compare_nbins}B and F
show that for both interpolations when choosing a low $n_{\mathrm{bins}}$, the mean of $\nabla U[x(t)]$,
$\langle \nabla U\rangle$, is far from zero. This is in contradiction to the fact that the simulation is stationary.
With $\langle \nabla U\rangle$ far from zero, both $C^{\nabla U x}(t)$ and $G(t)$ depend on the value of $n_{\mathrm{bins}}$.
When $n_{\mathrm{bins}}$ is increased, $\langle \nabla U\rangle$ converges to zero, and both interpolations
agree on the shape of $C^{\nabla U x}(t)$ in Fig.~\ref{fig:compare_nbins}A and E,
$G(t)$ in Fig.~\ref{fig:compare_nbins}D and H, and $\Gamma(t)$ in Fig.~\ref{fig:compare_nbins}C and G,
independent of $n_{\mathrm{bins}}$. This demonstrates that a consistent interpolation is possible.
For the main text and in Appendix section~\ref{si:velocities}, we use a linear interpolation with
$n_{\mathrm{bins}} = 600$.

\subsection{\label{si:velocities}Discretized Velocity}

\begin{figure}
	\centering
	\includegraphics[width=0.72\textwidth]{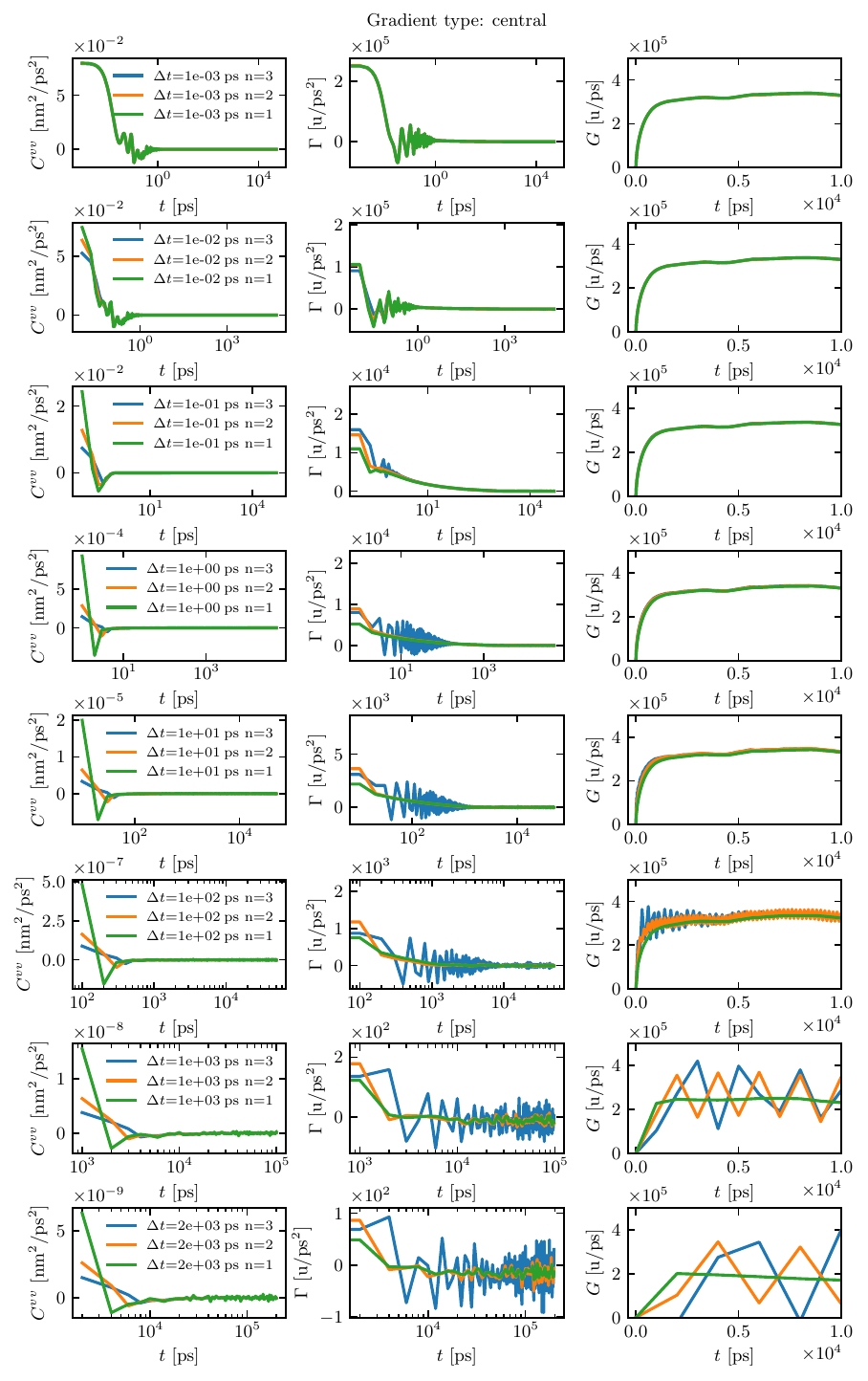}
	\caption{
		Memory extraction by the inversion of the Volterra equation~\ref{eq:volterra2} for different discretization times $\Delta t$, using MD data of Ala$_9$. Comparison of the memory kernels for central differences (Equation~\ref{eq:central_gradient}) of orders $n=1$ to $n=3$.
		The left column shows the velocity autocorrelation function $C^{vv}(t)$.
		The centre column shows the kernel $\Gamma(t)$. The right column shows the running integral over the kernel $G(t)$.
	}
	\label{fig:gradient_central}
\end{figure}

\begin{figure}
	\centering
	\includegraphics[width=0.72\textwidth]{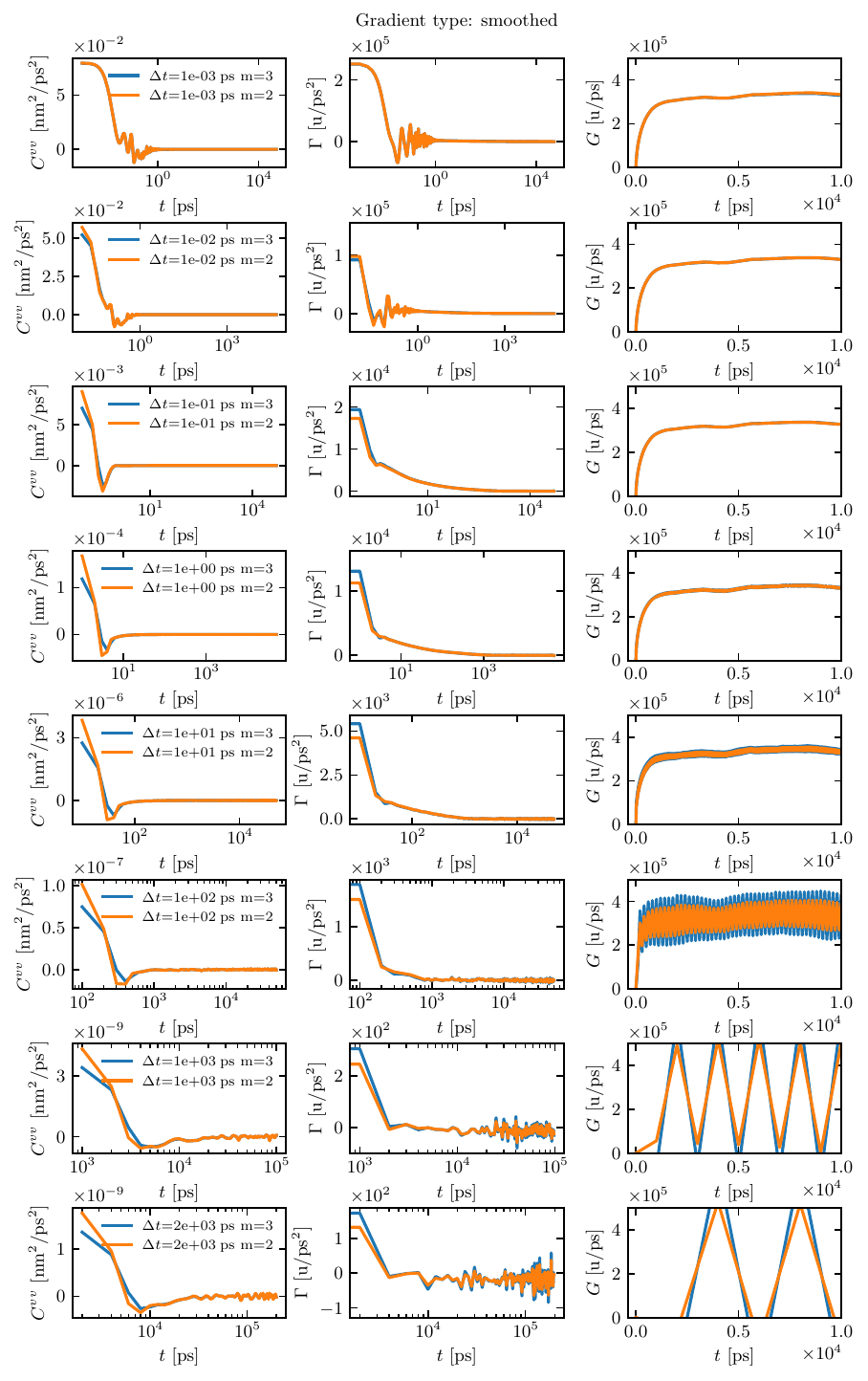}
	\caption{
		Memory extraction by the inversion of the Volterra equation~\ref{eq:volterra2} for different discretization times $\Delta t$, using MD data of Ala$_9$. Comparison of the memory kernels for smoothed differences (Equation~\ref{eq:smoothed_gradient})
		of orders $m=2,3$. The left column shows the velocity autocorrelation function $C^{vv}(t)$.
		The centre column shows the kernel $\Gamma(t)$. The right column shows the running integral over the kernel $G(t)$.
		We do not show order $m=1$, as it is equivalent to the central gradient of order $n=2$ shown in figure Fig.~\ref{fig:gradient_central}.
		(see Eq.~\ref{vel:equilivant}).
	}
	\label{fig:gradient_smoothed}
\end{figure}

To compute the velocities $v(t)$ from the positions $x(t)$, we compare two different differentiation schemes.
We define a central gradient of order~$n$:
\begin{equation}
	v_i^{(n)} = \frac{x_{i + n/2} - x_{i - n/2}}{n \Delta t}.
	\label{eq:central_gradient}
\end{equation}
Second, we define smoothed positions $\tilde{x}(t)$, where we smooth the data
by taking the running average over two adjacent points:
\begin{equation}
	\tilde{x}_i = \frac{x_{i + 1/2} - x_{i - 1/2}}{2}.
	\label{eq:smooth}
\end{equation}
We define smoothed positions of order $m$, where we apply the smoothing $m$ times
\begin{equation}
	\tilde{x}^{(m)}_i =
	\begin{cases}
		\tilde{x}_i & \text{if } m = 0\\
		(\tilde{x}^{(m-1)}_{i + 1/2} + \tilde{x}^{(m-1)}_{i - 1/2}) / 2 & \text{if} \; m \geq 1
	\end{cases}
	.
\end{equation}
The smoothed gradient of order $m$ is defined by smoothing the position $m$ times and then applying the central
gradient of order~1:
\begin{equation}
	\tilde{v}^{(m)}_i = \frac{\tilde{x}^{(m)}_{i + 1/2} - \tilde{x}^{(m)}_{i - 1/2}}{2}.
	\label{eq:smoothed_gradient}
\end{equation}
We note that the smoothed velocities of first-order equal the central gradient of second-order
\begin{equation}
	\begin{split}
		\tilde{v}_i^{(m = 1)} &= \frac{ \tilde{x}^{(m=1)}_{i + 1/2} - \tilde{x}^{(m=1)}_{i - 1/2}}{\Delta t}, \\
		&= \frac{ (x_{i + 1} + x_{i}) / 2 - (x_{i} + x_{i - 1}) / 2}{\Delta t}, \\
		&= \frac{ x_{i + 1} - x_{i - 1}}{2 \Delta t}, \\
		&= v_i^{(n = 2)}.
		\label{vel:equilivant}
	\end{split}
\end{equation}
Figures \ref{fig:gradient_central} and \ref{fig:gradient_smoothed} show that the
integrals over the kernel $G(t)$ that oscillate the least
are obtained by the central differences
of first order $v_i^{(n=1)}$. Accordingly, we use $v_i^{(n=1)}$ in the main text. In all cases, we compute the kernel $\Gamma(t)$ from $G(t)$ via a numerical gradient
using central differences of
second order, which alleviates the oscillations at low $\Delta t$.


\bibliography{main}

\end{document}